\documentclass[twocolumn,prb,showpacs,amsmath,amssymb,superscriptaddress,floatfix]{revtex4} 
\usepackage{graphicx}
\usepackage{dcolumn}
\usepackage{bm}
\usepackage{psfrag}

\begin{document}

\title{Phonon renormalization from local and transitive electron-lattice
couplings in strongly correlated systems}
 \author{E. von Oelsen}
 \affiliation{Institut f\"ur Physik, BTU Cottbus, PBox 101344,
          03013 Cottbus, Germany.}
 \author{A. Di Ciolo}
 \affiliation{SMC-INFM-CNR, Dipartimento di Fisica, Universit\`a di Roma
 ``La Sapienza'', P.  Aldo Moro 2, 00185 Roma, Italy.}
 \author{J. Lorenzana}
 \affiliation{SMC-INFM-CNR, Dipartimento di Fisica, Universit\`a di Roma
 ``La Sapienza'', P. Aldo Moro 2, 00185 Roma, Italy.}
 \affiliation{ISC-CNR, Via dei Taurini 19, 00185 Roma } 
 \author{G. Seibold}
 \affiliation{Institut f\"ur Physik, BTU Cottbus, PBox 101344,
          03013 Cottbus, Germany.}
 \author{M. Grilli}
 \affiliation{SMC-INFM-CNR, Dipartimento di Fisica, Universit\`a di Roma
 ``La Sapienza'', P.  Aldo Moro 2, 00185 Roma, Italy.}
\date{\today}
\begin{abstract}
Within the time-dependent Gutzwiller approximation (TDGA) applied to 
Holstein- and SSH-Hubbard models we study the influence
of electron correlations on the phonon self-energy.
For the local Holstein coupling we find that the phonon frequency
renormalization gets weakened upon increasing the onsite interaction
$U$ for all momenta. In contrast,  
correlations can enhance the phonon frequency shift 
for small wave-vectors in the SSH-Hubbard model.
Moreover the TDGA
applied to the latter model provides a mechanism which leads to 
phonon frequency corrections  at intermediate momenta due to the coupling 
with double occupancy fluctuations.
Both models display a shift of the nesting-induced 
to a $q=0$ instability when the onsite interaction becomes
sufficiently strong and thus establishing phase separation
as a generic phenomenon of strongly correlated electron-phonon
coupled systems.
\end{abstract}


\maketitle

\section{Introduction}
\label{sec:int}
Transition metal compounds are usually characterized 
by strong electron-electron and electron-phonon interactions 
(for an overview cf. 
Ref.~\onlinecite{maekawa}). The interplay of these interactions can
give rise to a large variety of interesting electronic properties
which also reflect in the energy and momentum structure of the phonons.
In this regard a well known phenomenon is the occurence of bond-stretching
phonon anomalies which are observed in high-T$_c$ cuprates 
La$_{2-x}$Sr$_x$CuO$_4$\cite{queeney99,pintch99} and YBa$_22$Cu$_3$O$_{6+x}$,\cite{Reich96} 
HgBa$_2$CuO$_{4+\delta}$,\cite{uchiyama04} 
Nd$_{1.86}$Ce$_{0.14}$CuO$_{4+\delta}$,\cite{DAs02} Bi$_2$Sr$_{1.6}$La$_{0.4}$Cu$_2$O$_{6+\delta}$\cite{graf08} but also in Pb- and K-doped BaBiO$_3$,~\cite{braden02} 
Sr$_2$RuO$_4$,~\cite{braden07} 
La$_{1.69}$Sr$_{0.31}$NiO$_4$~\cite{tran02} and
La$_{1-x}$Sr$_x$MnO$_3$.\cite{reich99}

The interpretation of such experiments requires an understanding of 
the renormalization of phonons in a strongly correlated electron system
which is the purpose of the present paper.
Such an analyis is usually based on Hubbard-type models where
the two most popular variants to implement 
an electron-lattice coupling 
are either to consider the dependency of onsite energies or 
of the hopping integrals as a function of some atomic coordinates.
In the first case, when one restricts to the interaction between electron
density and coordinate of the same lattice site, the coupling
is usually termed a Holstein or molecular crystal model.~\cite{hol59} In the second case, when the hopping
between nearest-neighbor sites is expanded in terms of the
positions of these sites the resulting electron-lattice coupling
is often named after Su, Schrieffer and Heeger (SSH) which
have used this type of interaction for the analysis of 
solitons in polyacetylene.~\cite{ssh}

Since lattice vibrations are partially screened by the electrons
the latter can have a profound influence on the effective dispersion
of the phonons. One example is the Kohn
anomaly \cite{kohn} caused by the abrupt
change of electronic screening near wave-vectors ${\bf q}$ which are twice
the Fermi momentum $k_F$ (nesting). 
Moreover, electronic correlations alter the electron dynamics of the system
and thus also affect the phonon dispersion.
In case of a Holstein coupling, and due to the fact that correlations 
reduce the charge correlations at $k_F$ in favor of an enhancement
of the $q=0$ response, it has been shown \cite{Gri94,CDCG95} that the formation of a
charge density wave (CDW) is suppressed in favor of a phase separation 
instability. For the phonons this has the immediate consequence that the
softening is shifted from the Kohn anomaly wave-vector $q=2 k_F$ to $q=0$.

Naturally the interplay between lattice and electronic degrees of freedom
has already been investigated
by means of several techniques. Some of the approaches, 
like quantum Monte Carlo, \cite{HS83,Hir83, HS84, Hir85, Ber95, srini, Hua03}
exact diagonalization, \cite{dob94,dob94epl,lor94b,stephan} and  
Dynamical Mean Field Theory (DMFT) \cite{FJ95, Capo04, Kol04-1, Kol04-2, Jeo04, San05, San06}
are intrinsically non-perturbative but are 
numerically challenging
and suffer some limitations (small lattice sizes, no momentum resolution, 
etc.).
On the other hand (semi)analytical approaches like variational ones, 
slave boson and large-N expansion
\cite{baer85,tes93,Gri94, Kel95, Zey96, alder97, caprara, Koc04, Cap04, Cit01, Cit05} deal with infinite 
systems, but within approximate treatments. 
It should be noted that most of the work has been done on
the Hubbard-Holstein model. The papers which explicitely deal with
an Hubbard-SSH model \cite{Hir83,srini,baer85,tes93,alder97,caprara} 
mainly focus on the interplay between correlations
and dimerization or superconductivity, respectively.

In this paper we want to study the correlation effects on the phonon
excitations for both Holstein and SSH models on the same footing.
To this aim we need a method which is not numerically very demanding, but
still provides a quantitatively acceptable treatment of the strongly 
correlated regime. In this regard we find the Gutzwiller approximation (GA)
supplemented with RPA-type fluctuations, the so-called time-dependent 
GA (TDGA), a good compromise. This technique corresponds to 
Vollhardt's Fermi liquid approach\cite{vol84} with the fluctuations 
extended to finite frequencies and momenta. Electron-phonon
interactions will be treated in the Born-Oppenheimer approximation.  

It is worth mentioning that the  TDGA approach
has been tested in various situations and found to be accurate
compared with exact diagonalization.~\cite{sei01,sei03,sei04b,sei07b} 
Computations for realistic models have provided a description of
different physical quantities in agreement with
experiment.~\cite{lor02b,lor03,sei05}

In the discussion of results we will restrict to one-dimensional 
systems. This clearly simplifies the computations and the presentation
of the results. It has the drawback that strictly one dimensional
systems are those for which our Fermi liquid like 
approach is expected to be less suited. Thus our results should not be
taken too literally in this case.  On the other hand the
qualitative behavior we find is rather independent of dimension. 
For example for the Holstein model we find here the same qualitative
behavior for the charge response as  we have found before in larger
dimensions.~\cite{diciolo08} Despite the inadequacy of a Fermi liquid
treatment for strictly one dimensional systems several real systems
are only quasi one dimensional and in those cases our computations can
be applicable. We also found that in certain filling ranges the
results are surprisingly accurate. 


The scheme of our paper is as follows.  
In Sec. II we define the model and show how the Hubbard-Holstein and
-SSH hamiltonians are represented within the GA.
After introducing the TDGA in Sec. IIC the phonon self-energy for both
Holstein- and SSH-couplings is derived in Sec. IID,E.
In Sec. IIIA we first investigate the quality of our technique by
comparing with exact and Monte-Carlo results in one dimension and
furtheron present the results for the phonon self-energies of the
Holstein- and SSH-Hubbard model in Secs. IIIB,C. 
Our conclusions can be found in Sect. VII and the details of our calculations are given in the Appendix. 

\section{Formalism}
\subsection{Model}
Our investigations are based on the following hamiltonian
\begin{equation}\label{eq:mod}
H = H_{e} + H_{e-ph} + H_{ph}
\end{equation}
where $H_{e}$ denotes the Hubbard model, $H_{e-ph}$ the coupling
between electrons and phonons and $H_{ph}$ the bare phonon part.
Here we restrict to one-dimensional systems and consider for the
electronic part hopping between nearest neighbors 
\begin{equation}\label{eq:hub}
H_{e}=-t\sum_{i,\sigma}\left(c_{i,\sigma}^\dagger c_{i+1,\sigma}
+ c_{i+1,\sigma}^\dagger c_{i,\sigma}\right) + U\sum_i n_{i,\uparrow}n_{i,\downarrow}
\end{equation}
where $c_{i,\sigma}^{(\dagger)}$ destroys (creates) an electron on lattice
site $R_i$ and $n_{i,\sigma}=c_{i,\sigma}^\dagger c_{i,\sigma}$.

We consider two types of electron-phonon coupling. The first is a
local Holstein interaction initially motivated from a molecular
crystal model 
\begin{equation}\label{eq:hol}
H_{e-ph}^{hol}=-\alpha \sum_{i,\sigma} u_i \left( n_{i,\sigma}- \langle 
n_{i,\sigma}\rangle\right)
\end{equation}
where $u_i$ is the coordinate of an internal mode of the molecules
affecting the site energy at site $i$. Its dynamics is described by
\begin{equation}\label{eq:hollat}
H_{ph}^{hol}=\frac{1}{2}K\sum_i u_i^2 +\frac{1}{2M}\sum_i p_i^2
\end{equation}
and corresponds to dispersionless lattice modes. Here $K$, $M$ denote
elastic constant and reduced mass and $p_i$ are the conjugate momenta 
at site $i$.

The second type of coupling originates from the dependence of the
electronic hopping on the atomic coordinates. For the
one-dimensional model under consideration this so-called
SSH or Peierls interaction is given by
\begin{equation}\label{eq:ssh}
H_{e-ph}^{ssh}= -\alpha t \sum_{i,\sigma}\left(u_{i+1}-u_i\right)
\left(c_{i,\sigma}^\dagger c_{i+1,\sigma}
+ c_{i+1,\sigma}^\dagger c_{i,\sigma}\right)
\end{equation}
and in this case the lattice dynamics is determined from
\begin{equation}\label{eq:phssh}
 H_{ph}^{ssh}=\frac{1}{2}K\sum_i \left(u_{i+1}-u_i\right)^2 
+\frac{1}{2M}\sum_i p_i^2 .
\end{equation}

\subsection{Gutzwiller approximation}
We treat the model Eq. (\ref{eq:mod}) within the Gutzwiller approximation (GA)
supplemented by Gaussian fluctuations the evaluation of which are outlined
in the next section. The GA can either be motivated from a slave-boson
approach \cite{Kot86} or a variational Ansatz formally evaluated
in infinite dimensions.~\cite{Geb90} 
The variational wave function is
given by  $|\Psi\rangle=\hat{P}|\phi\rangle$, where
the Gutzwiller projector $\hat{P}$ acts on the Slater determinant
$|\phi \rangle$. This approach  incorporates the correlation 
induced renormalization of the kinetic energy and treats the Hubbard on-site
interaction via the variational double occupancy parameters $D_i$. 
For each term in the Hamiltonian we derive an energy functional 
$E^{GA}[\rho,D]\equiv\langle\Psi|H|\Psi\rangle$
where we have introduced the one body density matrix associated
with the Slater determinant $\rho_{i,j,\sigma}=\langle\phi|
c_{i,\sigma}^\dagger c_{j,\sigma}|\phi\rangle$.
Specifically the energy functional for the
electronic part Eq. (\ref{eq:hub}) reads as
\begin{eqnarray}
E_{e}^{GA}[\rho,D]&=& - t\sum_{i,\sigma}z_{i,\sigma}z_{i+1,\sigma}\left(
\rho_{i,i+1,\sigma} + \rho_{i+1,i,\sigma}\right)\nonumber \\
&+& U\sum_i D_i\label{eq:hubga}
\end{eqnarray} 
and  the $z$-factors are given by
\begin{equation}\label{eq:hop}
z_{i\sigma}=\frac{\sqrt{(\rho_{ii,\sigma}-D_i)(1-\rho_{ii}+D_i)}+
\sqrt{(\rho_{ii,-\sigma}-D_i)D}_i}{\sqrt{\rho_{ii,\sigma}(1-\rho_{ii,\sigma})}}.
\end{equation}
Further on $\rho_{ij}=\sum_\sigma \rho_{ij,\sigma}$.
 
Within the Born-Oppenheimer approximation the electronic expectation 
value of $H_{e-ph}$ determines the lattice potential.   
In contrast to the local Holstein coupling, for electronic degrees
the transitive 
SSH electron-phonon interaction Eq. (\ref{eq:ssh}) is also renormalized by the 
z-factors and the corresponding energy functional reads as
\begin{eqnarray}
E_{e-ph}^{ssh,GA}&=& - \alpha t \sum_{i,\sigma}\left(u_{i+1}-u_i\right)
z_{i,\sigma}z_{i+1,\sigma}\nonumber \\
&\times& \left(\rho_{i,i+1,\sigma}+\rho_{i+1,i,\sigma}\right).
\label{eq:ssh2}
\end{eqnarray}

The GA variational ground state is then obtained
upon minimizing $E^{GA}$ with respect to $\{D\}$, $\{u\}$, and $\rho$ under the
constraint that the latter derives from a Slater determinant, i.e.
$\rho^2=\rho$. In the following our starting point will be an
homogeneous state. 
The formation of possible charge-density wave (in case of
the Holstein model) and dimerized (in case of the SSH model) states
will appear as instabilities of the homogeneous state. 
Thus our initial ground state is characterized by $u_i\equiv 0$, $\rho_{ii,\sigma}\equiv
\rho_0/2$, $D_i \equiv D_0$ and $z_{i}\equiv z_0$.
The ground state energy per site is therefore simply determined from 
Eq. (\ref{eq:hubga})
and reads
\begin{equation}
E^{GA}[\rho,D]/N = z_0^2 e_0 + U D_0
\end{equation}
and $e_0$ denotes the energy per site of a non-interacting
system with charge density $\rho_0$.
For later use we also denote the critical value of
the onsite repulsion $U=U_c = 32t/\pi$ for a half-filled 
one-dimensional
chain at which the Brinkman-Rice transition (i.e. complete
localization of the charge carriers with $D_0=0$) takes 
place.

\subsection{Time-dependent Gutzwiller approximation}\label{sec:ga}
In order to study fluctuations beyond the GA saddle-point solution,
necessary for the evaluation of the phonon self-energies,
we use the time-dependent GA (TDGA) which has been developed in
Refs.~\onlinecite{sei01,sei03}.

We briefly illustrate the formalism
for the electronic part $H_{e}$ and furtheron show how lattice fluctuations
can be implemented into the theory. Further details can be found in Refs.~\onlinecite{sei01,sei03,diciolo08}.

We study the response of the system to a small time dependent external
field which produces time dependent fluctuations in the density matrix
$\delta\rho$ and the double occupancy   $\delta D$.
This can then be obtained
by expanding $E^{GA}$ up to quadratic order in the density and
double occupancy fluctuations $\delta\rho$ and $\delta D$.

For a translationally invariant ground state it is convenient to
perform the expansion in momentum space.
Besides the local fluctuation $\delta \rho_{\bf q}$, we introduce the 
 bond charge fluctuation\cite{diciolo08},
$$\delta T_i=-t \sum_{\sigma\eta=\pm 1} (\delta
\rho_{i+\eta,i,\sigma} + \delta \rho_{i,i+\eta,\sigma}).$$
It is convenient to introduce the hopping factor in the definition of
$\delta T_i$ so that it can also be interpreted as a local kinetic
energy. Notice however that the $z$ factors are omitted.  
The Fourier transform is given by
\begin{equation}
\delta T_{q}=-2t\sum_{k,\sigma}\left[ \cos(k+q)+\cos(k) \right]\delta \rho_{k+q,k,\sigma}.
\label{Tq}
\end{equation}
For latter use it is also convenient to introduce the antisymmetric
combination of bond charge fluctuations: 
$$\delta T_i^-=-t \sum_{\sigma\eta=\pm 1} \eta (\delta
\rho_{i+\eta,i,\sigma} + \delta \rho_{i,i+\eta,\sigma}).$$
with Fourier transform:

\begin{equation}
\delta T_{q}^-=-2it\sum_{k,\sigma}\left[\sin(k+q)-\sin(k)\right]\delta
\rho_{k+q,k,\sigma}. 
\label{Tq-}
\end{equation}
It is easy to check that the two fluctuations are related by a
function of $q$:
\begin{equation}
  \label{eq:tvst}
\delta T_{q}^-=i \tan\left(\frac{q}2\right) \delta T_q  
\end{equation}
The second order-energy expansion in the charge channel of Eq. 
(\ref{eq:hubga}) follows as:
\begin{eqnarray}
E_{e}^{GA,(2)} &=& \frac{1}{N}\left[ \frac{1}{2}\sum_{q} V_q 
\delta \rho_q
 \delta \rho_{-q} 
+  z_0 z_{D}'\sum_{q}\delta D_q \delta T_{-q}
        \right. \nonumber \\
&+& \frac{1}{2} z_0(z'+z_{+-}')\sum_{q}
\delta \rho_{q}\delta T_{-q} 
 \nonumber \\
&+&\left.  \sum_q L_q \delta \rho_q \delta D_{-q}
+ \frac{1}{2} \sum_q U_q \delta D_q \delta D_{-q}\right]
\label{eskspace}
\end{eqnarray}
with the following definitions:
\begin{eqnarray*}
V_q &=& \frac{e_0 z_0}{2}(z_{++}''+ 2z_{+-}''+z_{--}'')
+\frac{1}{2}(z'+z_{+-}')^2e_0 \cos(q) \\
L_q &=& e_0z_0 (z_{+D}''+ z_{-D}'')
+z_D'(z'+z_{+-}')e_0 \cos(q) \\
U_q &=& 2e_0z_0 z_{D}''
+2(z_D')^2e_0 \cos(q)
\end{eqnarray*}
where $z'$ and $z''$ denote derivatives of the hopping 
factors which are given in the Appendix.

\begin{figure}[htbp]
\includegraphics[width=8cm,clip=true]{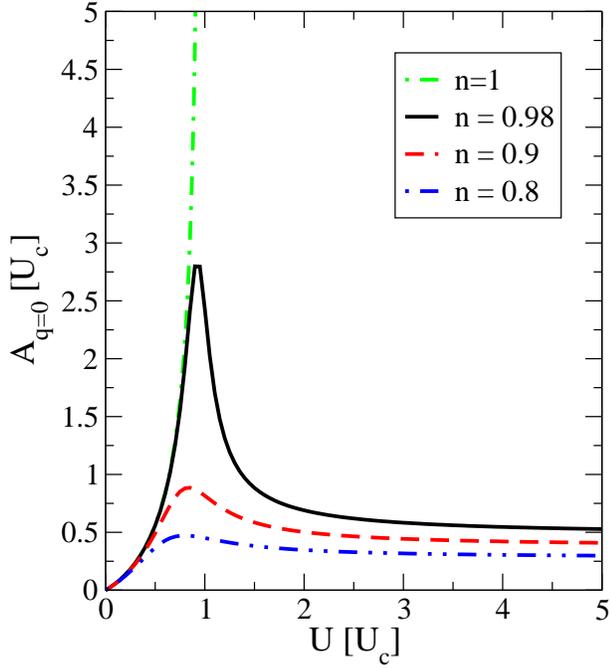}
\caption{(Color online) The interaction coefficient $A_{q=0}$ for various 
values of particle density $n$ in units of the Brinkman-Rice
critical onsite repulsion$U_c$ for a one-dimensional chain.}
\label{figaq}
\end{figure}

The double occupancy fluctuations can be expressed in terms
of the density fluctuations by use
of the antiadiabaticity condition which assumes that the double
occupancy adjust to the instantaneous configuration of the charge\cite{sei01}
\begin{equation}\label{eq:aac}
\frac{\partial E_{e}^{GA,(2)}}{\partial \delta D_q}=0
\end{equation}
and one obtains the following functional
which only depends on the local and intersite charge deviations:
\begin{equation}\label{eq:eexp}
E_{e}^{GA, (2)} =\frac{1}{2N}\sum_{q} \left(\begin{array}{c}
\delta \rho_q \\ \delta T_{q}\end{array}\right)
\left(\begin{array}{cc} A_q & B_q \\
B^*_q & C_q \\ 
\end{array}\right)
\left(\begin{array}{c}
\delta \rho_{-q} \\ \delta T_{-q}\end{array}\right)
\end{equation}where
\begin{eqnarray}
\underline{\underline{W_{q}}}^{e-e} =\left(\begin{array}{cc}
A_{q} & B_{q}\\
B^*_{q} & C_{q}\end{array}\right)
\label{Wqgen}
\end{eqnarray}
is the interaction kernel.
The elements of $\underline{\underline{W_{q}}}^{e-e}$ are given by
\begin{eqnarray}
A_q &=& V_q -\frac{L_q^2}{U_q} \label{eq:aq}\\
B_q &=& \frac{1}{2}
z_0 \left(z'+z_{+-}' -  z'_D
\frac{2L_q}{U_q}\right)\nonumber \\ 
C_q &=& - \ \  4\frac{(z_0z_D')^2}{U_q}\cos^2(q/2)\nonumber
\end{eqnarray}
and the long-wave limit of the $(11)$ element $A_ {q=0}$, which dominates the
interaction kernel close to half-filling is shown
in Fig. \ref{figaq}. For small $U$ one finds $A_{q=0}\approx U/2$
whereas $A_{q=0}$ is enhanced close to the Brinkmann-Rice transition 
$U=U_c$. At exactly half-filling one finds
\begin{equation}
A_{q} = \frac{U(U_c + U)(U-2U_c)}{4U_c(U -U_c)}.
\end{equation}
for $U<U_c$.

Since the energy expansion in Eq.~(\ref{eq:eexp}) is a quadratic form in
$\delta \rho_{q}$ and $\delta T_q$ (see also Eq.(\ref{Tq})), it is 
useful to introduce
the following susceptibility matrix for the non-interacting system
$\chi^{0}_{q}$:
\begin{widetext}
\begin{equation}
\label{sus}
\underline{\underline{\chi^{0}_{q}(\omega)}}=  \frac{1}{N} \sum_{k\sigma}\left(\begin{array}{cc}
1 & -2t\left[\cos(k) + \cos(k+q)\right]\\
-2t\left[\cos(k) + \cos(k+q)\right] & 
4t^2 \left[ \cos(k)+\cos(k+q)\right]^2 \end{array}\right)
 \frac{n_{k+q,\sigma} - n_{k\sigma}}
{\omega+\epsilon_{k+q}-\epsilon_{k}-i 0^+}.
\end{equation}
\end{widetext}
The susceptibility for the interacting system is then obtained
from the following RPA series 
\begin{equation}
\label{eqrpa}
\underline{\underline{\chi_{q}}} = \underline{\underline{\chi_q}}^{0} +
\underline{\underline{\chi_q}}^{0}\underline{\underline{W_{q}}}^{e-e}  
\underline{\underline{\chi_{q}}}
\end{equation}
where the element $(\chi_q)_{11}$ 
corresponds to the correlation function for the local charge response.

\subsection{Phonon self-energy for the Holstein coupling}
As mentioned above, due the local nature of the Holstein
coupling Eq. (\ref{eq:hol}) it is not renormalized by the
z-factors and thus its quadratic contribution to the
energy expansion is given by
\begin{equation}\label{eq:hol2}
E_{e-ph}^{hol,(2)}=-\alpha \frac{1}{N}\sum_{q} Q_{-q} \delta\rho_q
\end{equation}
where $Q_q=\sum_i \exp(-i q R_i) u_i$ denotes the Fourier 
transformed (normal) coordinate fluctuation (remember that the
saddle-point solution has $u_i=u_0=0$ so that we can skip the $\delta$
symbol).

Similar we write the lattice energy Eq. (\ref{eq:hollat}) as
\begin{equation}
E_{ph}^{hol,(2)}=\frac{1}{2N}\sum_q \left\lbrace \frac{P_qP_{-q}}{M}
+M \Omega^2 Q_q Q_{-q}\right\rbrace
\end{equation}
with $\Omega^2=K/M$.

The small time dependent 
deviation from the electronic ground state $\delta\rho_q$ will
act as a force on the lattice coordinates via $E_{e-ph}^{hol,(2)}$ and the
corresponding equation of motion reads
\begin{equation}
M\ddot{Q}_q +M\Omega_q^2 Q_q = -N\frac{\partial E_{e-ph}^{hol,(2)}}{\partial Q_{-q}}
=\alpha \delta\rho_q.
\end{equation}
As a consequence the lattice vibrations are shifted to new frequencies
$\omega_q$ which depend on the electronic charge fluctuation $\delta\rho_q$
\begin{equation}\label{eq:omeg1}
\left(\omega_q^2-\Omega^2\right) Q_q  = -\frac{\alpha}{M} \delta\rho_q .
\end{equation}
On the other hand $\delta\rho_q$ can be determined from linear response
theory when we view Eq. (\ref{eq:hol2}) as a small perturbation
on the electronic system. With the charge susceptibility derived
in the previous section one has
\begin{equation}
\delta\rho_q = (\chi_q)_{11} (-\alpha Q_q)
\end{equation}
which upon inserting in Eq. (\ref{eq:omeg1}) yields 
\begin{equation}\label{eq:elphonrenorm}
\omega_q^2=\Omega^2   + \frac{\alpha^2}{M} (\chi_q)_{11}
\equiv\Omega^2   + 2\Omega \Sigma_q
\end{equation}
where $\Sigma_q$ denotes the phonon self-energy.
In these equations $\chi_q(\omega)$ and $\Sigma_q(\omega)$ should be
evaluated at $\omega=\omega_q$. Since the phonon dynamics is assumed
to be much slower than the electron dynamics it is a good
approximation to evaluate the susceptibilities in the static limit.   

We see that in the case of the Holstein coupling the phonon frequency shift 
is solely determined  by the local charge susceptibility
renormalized by electronic correlations within the GA. 
For later use we define the ratio between self-energies in the
correlated and uncorrelated case
\begin{equation}\label{eq:gamma}
\Gamma_q \equiv \frac{\Sigma_q(U)}{\Sigma_q(U=0)}
\end{equation}
which for the Holstein coupling becomes
\begin{equation}
\Gamma_q = \frac{(\chi_q)_{11}}{(\chi^0_q)_{11}}.
\end{equation}
In addition it is convenient to introduce the following coupling constant
which has energy units
\begin{equation}
g_{hol}=\alpha\sqrt{\frac{1}{2M\Omega}}
\end{equation}
so that the phonon self-energy for the Holstein model is given by
\begin{equation} 
\Sigma_q=g^2_{hol}(\chi_q)_{11}.
\end{equation}

\subsection{Phonon self-energy for the SSH coupling}\label{secssh}
The case of the SSH electron-lattice coupling is more subtle since
the interaction energy Eq. (\ref{eq:ssh2}) depends on the
z-factors the fluctuations of which we have to consider in the
evaluation of $\Sigma_q$.

Expansion of Eq. (\ref{eq:ssh2}) up to second order in the
fluctuating fields yields for the Fourier transformed effective
electron-lattice interaction
\begin{eqnarray}\label{eq:ephssh}
E_{e-ph}^{ssh,(2)}&=&\alpha z_0^2 \frac{1}{N}\sum_q 
\delta T_q^- Q_{-q} \\
&+& i \alpha e_0 z_0 (z'+z_{+-}') \frac{1}{N}\sum_q \sin(q) \delta\rho_q Q_{-q}
\nonumber \\
&+& 2 i \alpha e_0 z_0 z_D' \frac{1}{N}\sum_q \sin(q) \delta D_q Q_{-q} 
\nonumber
\end{eqnarray}
with the same definitions already introduced in Sec. \ref{sec:ga}.
Besides the coupling to the transitive fluctuations $\delta T_q$ the
correlations induce a coupling of the lattice to local density 
($\delta\rho_q$) and double occupancy ($\delta D_q$) fluctuations.
Unlike the Holstein case the antiadiabaticity condition now includes the
electron-lattice coupling Eq. (\ref{eq:ephssh}) in addition to 
 the bare electronic part Eq. (\ref{eq:aac}),
\begin{equation}\label{eq:aacssh2}
\frac{\partial \left[ E_{e}^{GA,(2)} + E_{ph}^{ssh,(2)}\right]}{\partial \delta D_q}=0
\end{equation}
so that the double occupancy fluctuations can be expressed via the
density  {\it and} lattice fluctuations:
\begin{equation}\label{eq:dd}
\delta D_q = 2 i \alpha z_0 z_D' \frac{\sin(q)}{U_q}Q_q
-\frac{L_q}{U_q}\delta\rho_q -\frac{z_0z_D'}{U_q}\delta T_q.
\end{equation}

Inserting Eq. (\ref{eq:dd}) into Eqs. (\ref{eskspace},\ref{eq:ephssh})
and including also the (fourier transformed) lattice part 
Eq. (\ref{eq:phssh}) yields
\begin{equation}
E_{tot}^{ssh,(2)}=E_{e}^{GA, (2)}+E_{e-ph}^{ssh,(2)}+E_{ph}^{ssh,(2)}
\end{equation}
where $E_{e}^{GA, (2)}$ was derived in  Sec.\ref{sec:ga}, Eq. (\ref{eq:eexp}).
The effective coupling of the lattice to the electronic density
fluctuations is given by
\begin{equation}
E_{ph}^{ssh,(2)}=  \alpha\frac{1}{N}\sum_q W^1_q \delta\rho_q Q_{-q}+
\alpha \frac{1}{N}\sum_q  W^2_q \delta T_q Q_{-q} 
\end{equation}
where we used Eq.~(\ref{eq:tvst}) to eliminate the antisymmetric
fluctuations and introduced the elements of the vector
 ${\bf W_q}^{el-ph}$ 
\begin{eqnarray}
W_q^1 &=& i e_0 z_0\sin(q)\left[ z'+z_{+-}'  
-2 z_D'\frac{L_q}{U_q}\right] \label{eq:w1}\nonumber\\
&&\\
W_q^2&=&i z_0^2 \tan(q/2)
- 2 i e_0 (z_0 z_D')^2 \frac{\sin(q)}{U_q}.\nonumber  
\label{eq:w2}
\end{eqnarray}

The lattice part becomes
\begin{equation}
E_{ph}^{ssh,(2)}=\frac{1}{2N}\sum_q \left\lbrace \frac{P_qP_{-q}}{2M}
+M \Omega_q^2 Q_q Q_{-q}\right\rbrace .
\end{equation}
Interestingly the elimination of the double occupancy introduces a
novel renormalization of the phonon dispersion
\begin{equation}\label{eq:omeff}
\Omega_q^2=2 \frac{K}{M}\left(1-\cos(q)\right)
-\frac{1}{M}\left(2\alpha e_0 z_0 z_D'\right)^2 \frac{\sin^2(q)}{U_q}.
\end{equation}
The (squared) dispersion is composed of the acoustic branch $\sim 1-\cos(q)$
of the uncorrelated atomic chain and a contribution which arises
from the double occupancy fluctuations. To the best of our knowledge
this is the first time such correlated renormalization of the phonons
is reported. We will show below that it induces a phonon softening
in addition to the contribution which comes from electronic screening.

The phonon self-energy in the
present case can be derived via a similar procedure as before, 
taking into account the vectorial character of the susceptibility, and reads
\begin{equation}\label{eq:sigmassh}
\Sigma_q = g_q^2 \left[{\bf W_q}^{el-ph}\right]^T 
\underline{\underline{\chi_{q}}} {\bf W_q}^{el-ph}
\end{equation}
where the electronic susceptibility matrix $\underline{\underline{\chi_{q}}}$
is obtained from Eq. (\ref{eqrpa}). We also defined 
the dimensionless coupling constant
\begin{eqnarray}
g_q &=& \sqrt{\frac{\tilde{g}}{\hbar\Omega_q}}\label{eq:coup} \\
\tilde{g}&=& \alpha^2\frac{\hbar^2}{2M}
\end{eqnarray}

In the uncorrelated limit $U=0$ the self-energy reduces to
\begin{equation}
\Sigma_q(U=0) = 4 g_q^2(U=0) \sin^2(q/2) (\chi^0_q)_{22}
\end{equation}
where the coupling constant Eq. (\ref{eq:coup}) has to be
evaluated with the bare acoustic dispersion
$\Omega_q^2(U=0)=2 \frac{K}{M}\left(1-\cos(q)\right)$.

Finally it should be noted that the parameters of the problem
are given by $\hbar\omega_0 \equiv \sqrt{K/M}$, $\tilde{g}$ and $U$
measured in units of the hopping $t\equiv 1$ and we set $\hbar\equiv 1$
in the following.

\section{Results}
\subsection{Properties of the TDGA and comparison with exact results}
Since in this paper we restrict ourselves to one-dimensional systems
it is necessary to analyze the quality of our approximate TDGA scheme
in this special limit. With regard to the following investigation
of phonon self-energies it is important to figure out which features
can be considered as generic and also show up in higher dimensions
where our mean-field type approach is expected to perform better.

Since the phonon self-energies are completetly determined by the
charge susceptibility (at least in the Holstein case)
$(\chi_{q})_{11}$ we first analyze the corresponding TDGA result
which for small wave-vectors $q$ and close to half-filling can be expanded as
\begin{equation}\label{eq:chiq}
(\chi_q)_{11}\approx \frac{2}{\pi}\frac{v_F q^2}{\omega^2-(v_{\rho}q)^2}
\end{equation}
where $v_F=v_F^0z_o^2$ is the quasiparticle Fermi velocity and
$v_{\rho}=v_F\sqrt{1+4A_0/(\pi v_F)}$ is the velocity of
the (quasi)particle-hole excitations with $A_0\equiv A_{q=0}$ defined
in Eq. (\ref{eq:aq}).
The compressibility $\kappa = -(\chi_{q\to 0})_{11}(\omega=0)$
follows as
\begin{equation} 
\kappa = \frac{2v_F}{\pi v_\rho^2}
=\frac{2}{\pi}\frac{1}{v_F+4 A_0/\pi}
\end{equation}
In the weak coupling limit these expression coincide with the
perturbative expressions for the Tomonoga-Luttinger
liquid.~\cite{voit} As in the exact case the
effective interaction and the Fermi velocity gets renormalized upon
increasing $U$.

\begin{figure}[htbp]
\centering
\includegraphics[width=8.cm,clip=true]{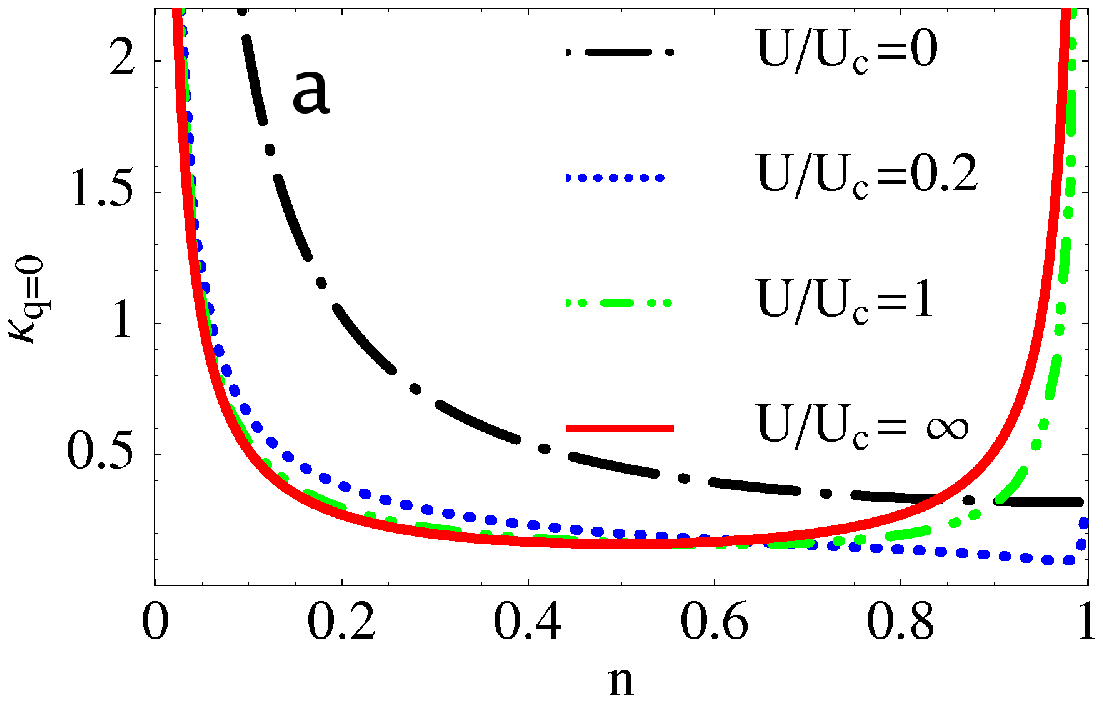}\hspace{.1 cm}%
\includegraphics[width=8.cm,clip=true]{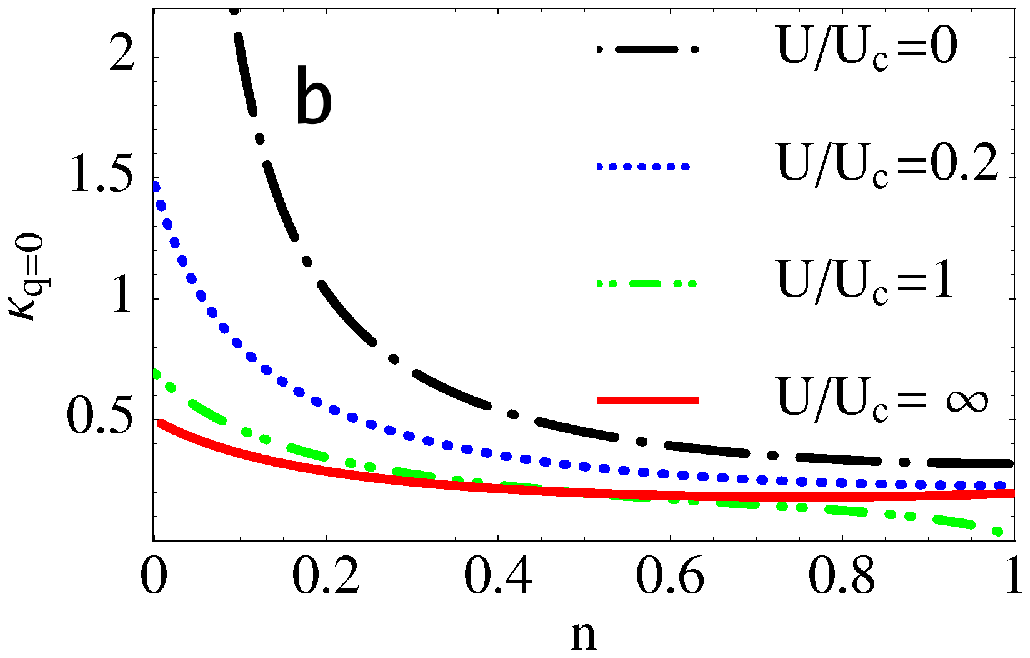}
\caption{(Color online) 1d-Charge compressibility as a function of $n$ and for different values of $U/U_c$ calculated with the Bethe ansatz - 
exact 1d solution (a) and with the TDGA (b).
Here $U_c = 32t/\pi$ is the Coulomb repulsion 
at which the Brinkman-Rice transition takes place for $n=1$ in the GA. 
In the exact solution the metal insulator transition occurs at $U=0$ 
therefore $U_c= 32t/\pi$ is used only as an energy unit.
}
\label{mykSchulz}
\end{figure}

In Fig.~\ref{mykSchulz} we compare the TDGA charge compressibility
with the exact  1d solution of the Hubbard model.~\cite{LW68, Shi72,Sch90}
The renormalization of $v_F$ and the effective interaction  
pushes the qualitative agreement with exact results to larger
values of $U$ than the traditional HF+RPA approach.
Strong differences arise close to
$n=1$. As soon as the interaction is switched on the exact compressibility
diverges. This can be understood in the strong coupling limit where the
charge degrees of freedom  can be mapped to a spinless Fermion
model\cite{oga90} and the compressibility is related to the spinless
density of states which has a 1d van Hove divergence. In contrast the GA yields
a compressibility which tends to zero at the BR point. 
We remark that the GA compressibility has a jump discontinuity 
for $n=1$ and $U>U_c$. 
In fact, its left and right limits are finite, while its value computed 
in $n=1$ is zero. At half-filling  
an antiferromagnetic  (AF) broken symmetry TDGA computation instead of the
present paramagnetic one yields much more accurate results.~\cite{sei01}

In the dilute
limit the exact compressibility diverges again whereas the TDGA
result yields a finite value. This disagreement is not surprising
since the RPA in general is well known to fail at low densities. In this case a
particle-particle approach, recently implemented on top of the GA,~\cite{sei07b}
would be more appropriate.  

Despite the (expected) failure  of the paramagnetic TDGA at low and
half filling at intermediate fillings the behavior of the
compressibility as a function of $U$ is qualitative and to some extent
quantitatively reproduced. One should keep in mind again that we are using a
Fermi liquid approach whereas the real ground state is a Luttinger
liquid.  

\begin{figure}[htb]
\centering
\includegraphics[width=8cm,clip=true]{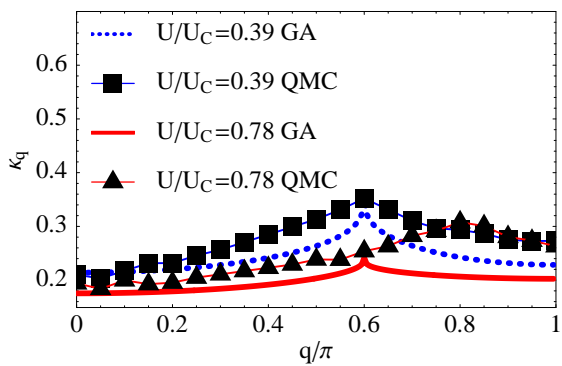}
\caption{(Color online) 1d-Charge susceptibility ($n=0.6$) as a function of $q/\pi$: comparison between TDGA and 
QMC results.~\cite{HS84}}
\label{mykHirsch}
\end{figure}

In Fig.~\ref{mykHirsch} we compare the TDGA charge susceptibility with 
QMC results from Hirsch and Scalapino.~\cite{HS83,HS84} 
Since their data are for $n=0.6$ we expect 
the TDGA to give reasonable results. Although our formalism is at $T=0$ and the 
QMC study at $T \neq 0$, the comparison is meaningful because their results 
are at $T = 0.0690$, quite low if compared to the electronic energy scales. 
The QMC susceptibilities generally agree with ours within ten-twenty percent 
deviations, and with larger deviations at high 
$q$ for $U \sim U_c$. For large $U$, in fact, the QMC data exhibit the 
transfer of the peak from $q=2k_F$ to $q=4k_F$, signature of the spin-charge 
separation of the 1d-Luttinger liquid, clearly absent in our 1d FL. 
The QMC curves present a finite $T$ effect 
that smoothes the peak. 
\begin{figure}[tp]
\centering
\includegraphics[width=8. cm,clip=true]{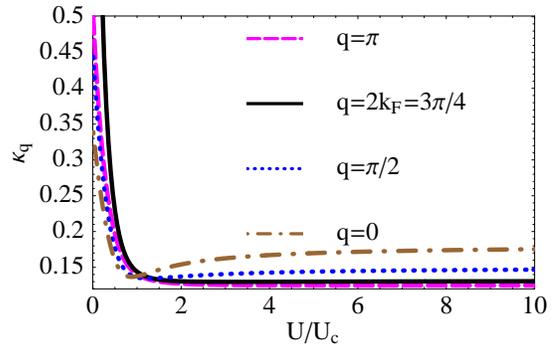}\hspace{.1 cm}
\caption{(Color online) 1d-Charge susceptibility as a function of $U/U_c$ for $n=0.75$.}
\label{fig:kdnp198q}
\end{figure}

The above results indicate that our FL scheme works quantitatively rather 
well away from half-filling (where an AF TDGA computation would do a 
better job) and provides reasonable momentum 
dependencies (but for the subtle Luttinger $4k_F$ peak shift for large $U$). 

In Fig.~\ref{fig:kdnp198q}, we show 
the charge susceptibility $\kappa_q$ as a function of $U/U_c$. These 
results should be taken with a pinch of salt due to the explained
drawbacks, however they illustrate well the general behavior of the TDGA
that are found in higher dimension where the approach performs
better.~\cite{diciolo08}   
For small deviations of the density 
from half-filling the compressibility has a minimum close to $U=U_c$
which is due to the corresponding maximum in the $(11)$ element
of the interaction kernel (cf. Fig. \ref{figaq}).  
For large $q$ this minimum 
becomes too shallow to be clearly seen in Fig.~\ref{fig:kdnp198q}. 
At small momenta the charge susceptibility is close to the compressibility. 
As the momentum approaches $q=2k_F=n\pi$ the charge susceptibility
diverges for small $U$. However, this divergence is strongly suppressed 
upon increasing $U$. 
At small doping $\kappa_q$ is finite but still shows a shallow minimum 
close to $U=U_c$. This behaviour is due to the proximity 
of the Mott phase which is  more clear in higher dimensions.~\cite{diciolo08}
The essential point is that the maximum charge response changes from
the wave-vector $q=2k_F$ to $q=0$ upon increasing $U$.
Therefore correlations suppress the nesting induced transition to
a CDW state in favor of phase separation as is discussed in more
detailed and higher dimensional systems in Ref.~\onlinecite{diciolo08}.

It is important to notice that the failures of the TDGA  found at
high filling  in our comparison with the exact 
results can be traced back to specific features of the 1d physics (like, e.g.,
the spin-charge separation, the equivalence to spinless fermions at 
$U=\infty$). Therefore these failures should not be attributed to the TDGA 
{\it per se}, but to the underlying 
assumption of a FL ground state.
This is why our results not only provide qualitative informations on the 
1D case, but also shed light on the physics of the FL state occurring at higher dimensions.

\subsection{Holstein Coupling}
\begin{figure}[htb]
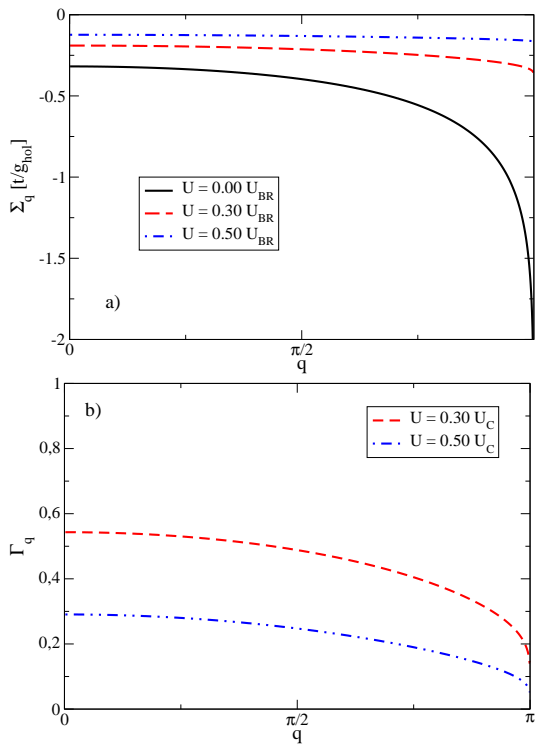

\centering
\includegraphics[width=7cm,clip=true]{FIG_5a.eps}
\includegraphics[width=7cm,clip=true]{FIG_5b.eps}
\caption{(Color online) a) Self-energy $\Sigma_q$ in units of $t/g_{hol}$ and
b) ratio between correlated and uncorrelated self-energies $\Gamma_q$
for the half-filled Holstein-Hubbard model as a function of momentum $q$.}
\label{1DGammaq}
\end{figure}

The (static) phonon self-energy for the Holstein coupling 
$\Sigma^{hol}_q\equiv \Sigma_q^{hol}(\omega=0)=g_{hol}^2[\chi_q]_{11}(\omega=0)$ 
corresponds to the local
charge correlation so that the discussion of the previous section 
directly applies
also here. Fig.~\ref{1DGammaq}a shows $\Sigma_q^{hol}$ for the half-filled
system where it is given by 
\begin{displaymath}
\Sigma_q^{hol}= g_{hol}^2\frac{[\chi^{0}_{q}]_{11}}{1-A_{q}[\chi^{0}_{q}]_{11}}
\end{displaymath}
As is clear from the previous section the primary purpose of these results 
is to illustrate the generic behavior
expected in higher dimensions rather than to comprise the physics of
1D systems.

Due to the divergency of  $[\chi^{0}_{q}]_{11}$ at $q=2k_F=\pi$ the
corresponding singularity in $\Sigma_q^{hol}$ gets suppressed upon
increasing $U$. This suppression persists for all momenta which
can also be seen from  the vertex $\Gamma_q$ (Fig.~\ref{1DGammaq}b,
cf. Eq. (\ref{eq:gamma}) which  quantifies the phonon frequency
shift for the correlated system as compared to the non-interacting
case:
\begin{equation}
\Gamma_q= \frac{1}{1-A_q[\chi^{0}_{q}]_{11}}.
\end{equation}
In the limit $q\to 0$ where $(-\chi^{0}_{q})_{11}$ equals the
density of states $N(E_F)$ the reduction of $\Gamma_{q=0}$ is therefore
determined  by the effective interaction $A_q$ which diverges
for $U$ approaching the Brinkmann-Rice transition (and thus $\Gamma_q\to 0$).
On the other hand, at $q=\pi$ the local non-interacting 
charge correlations $(\chi^{0}_{q=\pi})_{11}$ display a divergence due to
nesting and are thus responsible for the vanishing of $\Gamma_{q\to\pi}$
in this limit. 

\begin{figure}[htb]
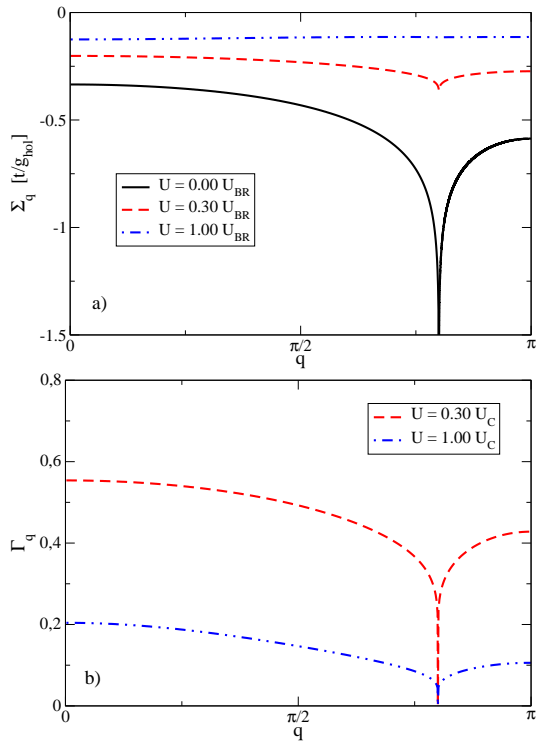

\centering
\includegraphics[width=7cm,clip=true]{FIG_6a.eps}
\includegraphics[width=7cm,clip=true]{FIG_6b.eps}
\caption{(Color online) a) Self-energy $\Sigma_q$ in units of $t/g_{hol}$ and
b) ratio between correlated and uncorrelated self-energies $\Gamma_q$
for the Holstein-Hubbard model (particle density $n=0.8$) 
as a function of momentum $q$.}
\label{1DGammaq2}
\end{figure}

At finite doping (Fig. \ref{1DGammaq2}a) the singularity of  
$\Sigma_q^{hol}$ at $U=0$ occurs at $q=2k_F<\pi$ and similar to
the half-filled system becomes suppressed upon increasing $U$.
As discussed in the previous section (cf. Fig.~\ref{fig:kdnp198q}), 
for large onsite interaction $U$ 
the charge susceptibility (and thus $\Sigma_q^{hol}$) acquires a maximum
at small momenta so that the dominant phonon renormalization is shifted
from $q=2k_F$ to $q=0$.
On the other hand (Fig. \ref{1DGammaq2}b) the
reduction of $\Gamma_q$
is still most pronounced at the Fermi momenta $q=2 k_F$
where the bare Lindhard susceptibility $(\chi^{0}_{q})_{11}$ logarithmically 
diverges whereas $\Sigma_{q=2k_F}(U>0)$ stays finite and thus
$\Gamma_{q=2k_F} =0$.
Of course this is peculiar to the one-dimensional
system where one has perfect nesting for each carrier density.

We now analyze in more detail the mode which becomes 
soft when the dominating instability shifts from $q=2 k_F$
to $q=0$ for large $U$ and close to half-filling.
The dressed phonon propagator
\begin{equation}
D_q(\omega)=\frac{D^0_q(\omega)}{1-g_{hol}^2[\chi_{q}]_{11}(\omega)
  D^0_q(\omega)}
\end{equation}
couples the Holstein phonon $\Omega_0$ with the energy of the 
particle-hole excitations $\sim v_{\rho}q$ 
when we use the long wavelength limit for the charge susceptibility
given in Eq. (\ref{eq:chiq}).
Then $D_q(\omega)$ acquires new poles at
\begin{eqnarray}
\omega^2_{+}&=& \Omega_0^2 + \frac{4g_{hol}^2 v_F}{\pi\Omega_0}q^2 \\
\omega^2_{-}&=& (v_\rho q)^2 - \frac{4g_{hol}^2 v_F}{\pi\Omega_0}q^2
\end{eqnarray}
corresponding to a hardening of the Holstein phonon and
a softening of the effective ('zero sound') particle-hole velocity.
Thus the instability at $q=0$ does not follow from a zero in
the phonon-type mode but due to the fact that the particle-hole
excitation acquire a negative velocity.
Since the poles of the dressed charge propagator are
identical to those of $D_q(\omega)$ this also corresponds
to a phase separation instability so that the present
approach generalizes the analysis of Ref.~\onlinecite{Gri94}
for $U\to \infty$ Hubbard models to finite onsite interactions.

\subsection{SSH coupling}
\begin{figure}[htb]
\centering
\includegraphics[width=8cm,clip=true]{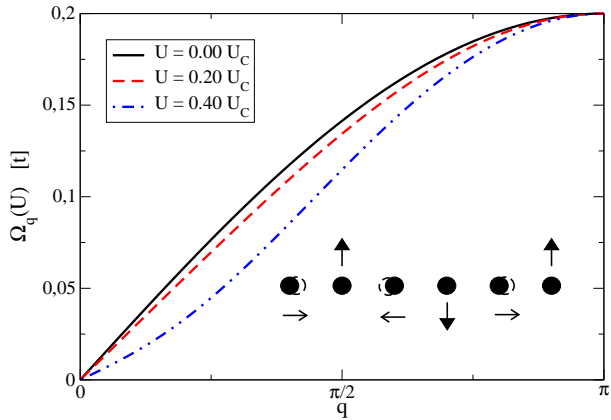}
\caption{(Color online) Acoustic phonon dispersion  
in the SSH model for three different 
$U$ values at half-filling ($\tilde{g}=0.01$, $\omega_0=0.1$). 
Shown is only the influence of the double occupancy 
fluctuations on $\Omega_q$. The inset displays atomic displacements with
wave-vector $q=\pi/2$ (horizontal arrows) and the associated modulation
of the double occupancy (vertical arrows) with the same periodicity.}
\label{omega}
\end{figure}

For the transitive electron-phonon coupling we have seen in Sec. \ref{secssh}
that correlations already induce a renormalization of the phonon
dispersion Eq. (\ref{eq:omeff}) 
\begin{equation}
\Omega_q = \sqrt{(\Omega_q^0)^2 - (\Delta\Omega_q)^2} 
\end{equation}
due to the elimination of the double occupancy
fluctuations (cf. Eqs. (\ref{eq:aacssh2},\ref{eq:dd})).
Here $\Omega_q^0=2\omega_0 \sin(q/2)$ denotes the acoustic branch for $U=0$ and
the correlation induced contribution $\sim -(\Delta\Omega_q)^2$ 
is always negative (since $U_q>0$ in Eq. (\ref{eq:omeff})). 
The corresponding softening of $\Omega_q$ is shown in Fig. \ref{omega}.

The renormalization vanishes for both $q\to 0$ and $q\to \pi$, and is
largest for intermediate momenta $q\sim \pi/2$.
This can be understood from Eq. (\ref{eq:dd}) where the
first term links the displacements to the double occupancy fluctuations
$\delta D_q \sim \sin(q)/U_q Q_q$. Remember that $U_q$ is the
interaction energy of double occupance fluctuations (cf. Eq. (\ref{eq:hubga})
which has a significant momentum dependence only close to half-filling
and large $U$. Therefore the spatial relation between $Q_q$ and $\delta D_q$
is mainly determined by  $\sin(q)$ and thus largest at $q\approx \pi/2$.
The inset to Fig. \ref{omega} depicts the corresponding lattice
modulation (horizontal arrows) which, due to the inreased (decreased)
hybridization, favors a modulation of the density and double occupancies
with the same periodicity (vertical arrows).

\begin{figure}[htb]
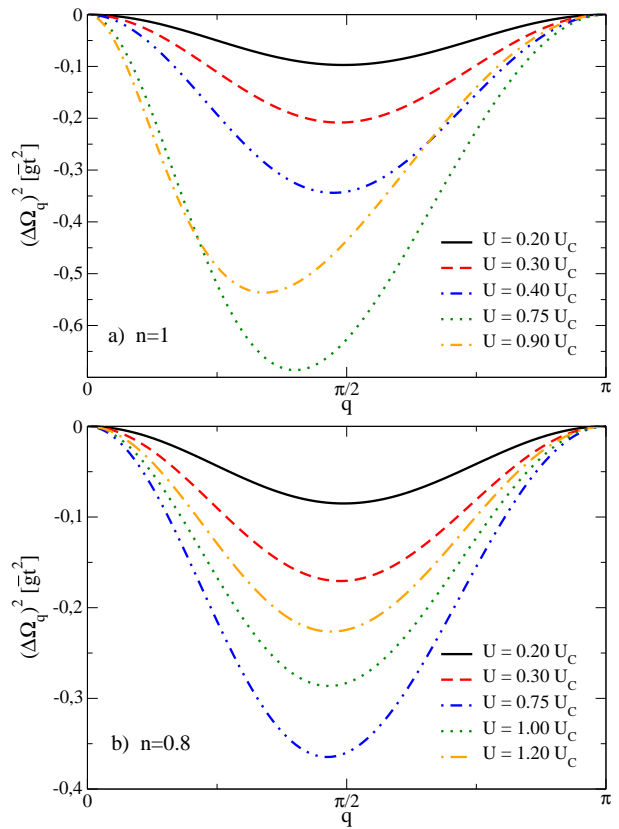

\centering
\includegraphics[width=8cm,clip=true]{FIG_8a.eps}
\includegraphics[width=8cm,clip=true]{FIG_8b.eps}
\caption{(Color online) Correlation induced correction to the phonon dispersion 
(in units of $\tilde{g}t^2$) 
from the elimination of double occupancy fluctuations.
Charge densities
are $n=1$ (panel a) and $n=0.8$ (panel b)}
\label{deltaomega}
\end{figure}
 
The correction $(\Delta\Omega_q)^2$ to the phonon dispersion induced by
the double occupancy fluctuations is separately displayed in 
Fig. \ref{deltaomega}. For the half-filled system (top panel) the
maximum of $(\Delta\Omega_q)^2$ shifts to smaller $q$-values upon
increasing $U$ due to the more significant momentum dependence
of $U_q$ as mentioned above. This is less pronounced for the
doped system (lower panel) where the maximum in the correlation
induced correction stays close to $q=\pi/2$. Note also that
$(\Delta\Omega_q)^2$ has a maximum as a function of $U$. This
is due to the fact that the SSH electron-phonon interaction
Eq. (\ref{eq:ssh2}) is renormalized by the $z$-factors which decrease
with increasing $U$ so that the transitive fluctuations become
suppressed. In this regard $(\Delta\Omega_q)^2$
results from a subtle interplay of kinetic and correlation effects.

We now turn to the influcence of electronic density fluctuations
on the phonon dispersion which is measured in terms of
the phonon self-energy $\Sigma_q$ Eq. (\ref{eq:sigmassh}). 

\begin{figure}[htb]
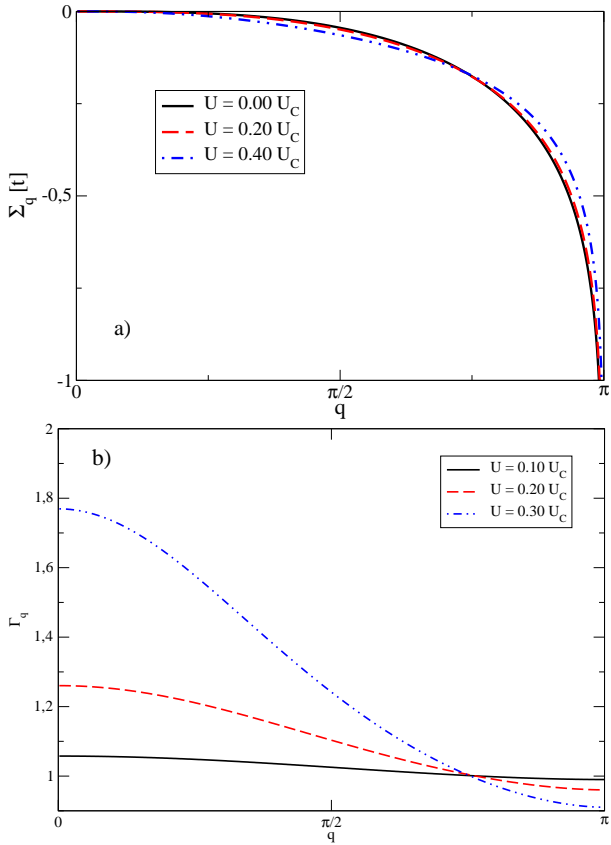

\centering
\includegraphics[width=8cm,clip=true]{FIG_9a.eps}
\includegraphics[width=8cm,clip=true]{FIG_9b.eps}
\caption{(Color online) a) Phonon self-energy $\Sigma_q$ for the half-filled 
Hubbard-SSH model and
different values of $U/U_{c}$. b) Ratio $\Gamma_q=\Sigma_q(U)/\Sigma_q(U=0)$ 
for the same system.
Parameters: $\tilde{g}=0.01$, $\omega_0=0.1$.}
\label{sigmassh}
\end{figure}

Fig. \ref{sigmassh}a displays $\Sigma_q$ for the half-filled system.
In this limit the local density fluctuations (originating from the
Hubbard interaction) are decoupled from the transitive ones. 
Therefore the latter  are not
screened and the divergence at $q=\pi$ in case of the SSH 
coupling is not removed upon 
increasing $U$ in contrast to the Holstein-Hubbard model. 
As a consequence, the
phonon excitations always 
(i.e. for infinitesimally small electron-phonon coupling)  
become instable for $q=\pi$ corresponding to the homogeneous 
dimerized state. Correlations lead to a suppression (enhancement) of
$\Sigma_q$ for large (small) momenta as can be more clearly seen from
Fig. \ref{sigmassh}b which shows the ratio 
$\Gamma_q=\Sigma_q(U)/\Sigma_q(U=0)$. In the limit $q=\pi$ one can
show that the ratio is given by $\Gamma_{q=\pi}=z_0^2$, i.e. it is
completely determined by the hopping renormalization factors of the GA.
This is consistent with the fact, that for the half-filled dimerized system 
correlations suppress the dimerization order parameter \cite{Hir83,baer85}
due to the
reduction of the effective electron-phonon coupling. 

In the limit $q\to 0$ the self-energy vanishes, however, the slope
of $\Sigma_{q\to 0}$ strongly depends on the correlations and leads
to the observed increase of $\Gamma_{q\to 0}$ with increasing $U$ (Fig.
\ref{sigmassh}b). The main reason for this enhancement comes
from the fact that the SSH coupling is a coupling to
transitive electronic correlations which at half-filling and small
momenta are decoupled from the local ones. In the
limit $q\to 0$ and half-filling Eq. (\ref{eq:sigmassh}) becomes
\begin{eqnarray}
\Sigma_q^{SSH} &\approx & g_q^2 |W^2_q|^2 (\chi_q^0)_{22} \\
W_q^2 &\approx& i q \left[ z_0^2 + 2\frac{U^2}{U_c^2}\right] = 
 i q \left[ 1 + \frac{U^2}{U_c^2}\right]\label{eq:vert}
\end{eqnarray}
and one finds that $\Sigma_q^{SSH}$ is not screened by the strong local
charge fluctuations. On the contrary, since $(\chi_q^0)_{22} \sim 1/z_0^2$
it becomes enhanced due to the increase of the
quasiparticle mass with $U$. However, for the bare SSH coupling
$|W^2_q|^2 = q^2 z_0^4$ this effect would be overcompensated resulting in 
$\Sigma_q \sim z_0^2$.  It is due to the TDGA induced vertex corrections
Eq. (\ref{eq:vert}) that the increase of $(\chi_q^0)_{22}$ is even amplified
by the concomitant increase of $|W^2_q|^2$ with $U$.
Finally, another (though much weaker) factor which leads to the enhancement
of $\Sigma_{q\to 0}^{SSH}$ with $U$ comes from the dependence of
the coupling constant $g_q$ Eq. (\ref{eq:coup}) on the phonon frequencies 
$g_q \sim 1/\sqrt{\Omega_q}$ which become softened due
to the elimination of the double occupancy fluctuations 
(cf. Eq. (\ref{eq:omeff})).

 The enhancement
of the vertex $\Gamma_q$ at small momentum is a new effect very much
in contrast with the result in the Holstein case\cite{diciolo08}
 where one always finds $\Gamma_q<1$, i.e.
a reduction of self-energy corrections with $U$.

\begin{figure}[htb]
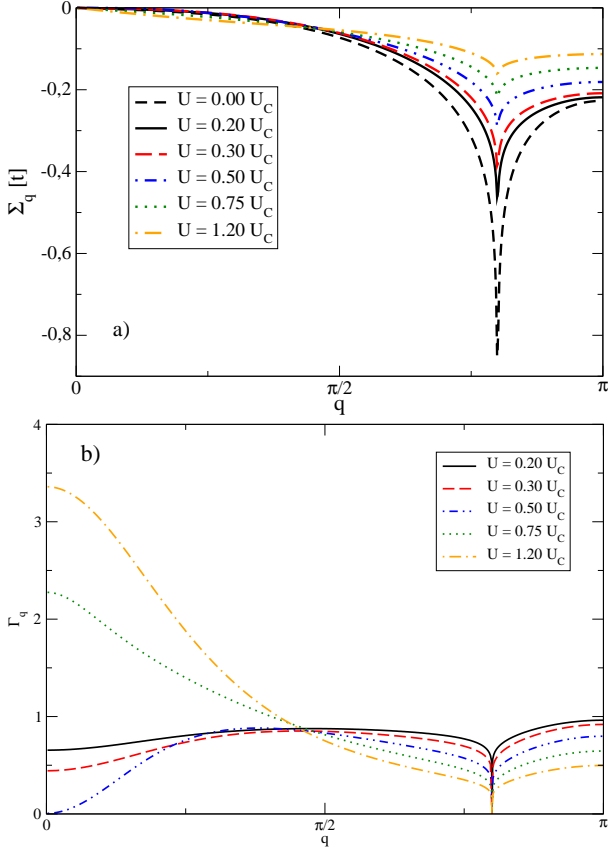

\centering
\includegraphics[width=8cm,clip=true]{FIG_10a.eps}
\includegraphics[width=8cm,clip=true]{FIG_10b.eps}
\caption{(Color online) a) Phonon self-energy $\Sigma_q$ for the doped ($n=0.8$) 
Hubbard-SSH model and
different values of $U/U_{C}$. b) Ratio $\Gamma_q=\Sigma_q(U)/\Sigma_q(U=0)$ 
for the same system.
Parameters: $\tilde{g}=0.01$, $\omega_0=0.1$.}
\label{gammassh}
\end{figure}

Fig. \ref{gammassh} displays  the behavior of $\Sigma_q$ and $\Gamma_q$
for the doped SSH model.
Similar to the case of half-filling, $\Sigma_q$ is reduced upon increasing
$U$ for large momenta. However, the behavior for small $q$ becomes
more subtle as can be seen from Fig. \ref{gammassh}b.
As a function of $U$ the self-energy $\Sigma_{q\to 0}$ 
passes through a minimum and for large $U$ exceeds again the uncorrelated
value (i.e. $\Gamma_{q\to 0} >1$)  similar to the half-filled case.
This behavior results from a subtle interplay between
local and transitive charge fluctuations which are now coupled.
For small $U$ the screening induced by the local charge fluctuations
leads to a suppression of $(\chi_q)_{22}$ and also $\Sigma_q$.
Only at larger $U$ the vertex corrections for  $|W^2_q|$ can overcome
this decrease and effectively enhance again the self-energy at small 
momenta.

The coupling of the local charge
density fluctuations also contributes to the suppression of the
$q=2k_F$ divergence in the self-energy.
As in case of the Holstein coupling one therefore finds 
that $\Gamma_{q=2k_F} =0$ since
$\Sigma_{q=2k_F}(U=0)$ logarithmically diverges 
whereas  $\Sigma_{q=2k_F}(U>0)$ stays finite. 

\begin{figure}[htb]
\centering
\includegraphics[width=8cm,clip=true]{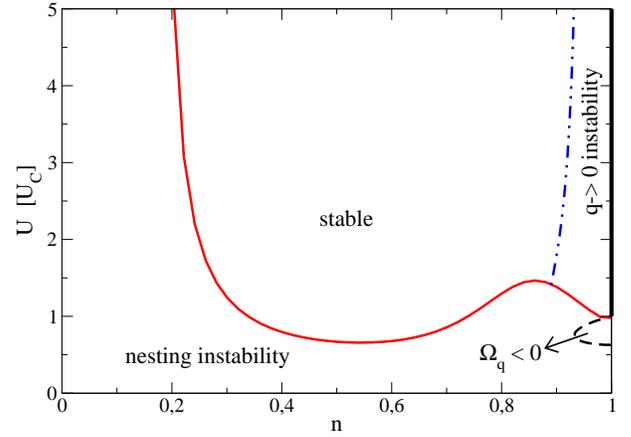}
\caption{(Color online) Phase diagram for the SSH model with
parameters $\tilde{g}=0.02$, $\omega_0=0.2$. The solid
line indicates the transition to a nesting induced $q=2k_F$
instability whereas the dashed-dotted line is the $q=0$ instability line.
The region enclosed by the dashed line corresponds to the
parameter space where $\Omega_q<0$ and the bar at $n=1$ 
indicates the localized regime for $U>U_{c}$.}
\label{phase}
\end{figure}

Within the Holstein-Hubbard model we have seen that away from half-filling
the maximum self-energy shifts from the nesting vector $q=2k_F$ to
$q=0$ when the correlations become sufficiently strong. As a consequence,
a large electron-phonon coupling will induce a CDW instability for
small, but a phase separation instability for large $U$, also
of course depending on the carrier density.
Is there a similar scenario for the transitive coupling in the SSH model?
We determine the instabilities from the zero
frequency poles of the phonon propagator
\begin{equation}
D_q(\omega=0)=\frac{D^0_q(\omega=0)}{1-\Sigma_q D^0_q(\omega=0)}
\end{equation}
which yields the condition 
\begin{equation}\label{eq:cond}
\Omega_q=-2\Sigma_q
\end{equation}
and $\Omega_q$ is the effective phonon dispersion given in 
Eq. (\ref{eq:omeff}). 
The solid line in Fig. \ref{phase} marks the lattice instability 
at $q=2k_F$, i.e. where the system undergoes a
transition towards a combined CDW and bond-order state. This instability
is suppressed for large $U$ due to the suppression of the $2k_F$ peak
in $\Sigma_q$ as shown in Fig. \ref{gammassh}.

The maximum at $n=0.85$ in the instability line is due the
following. At $n=1$ the self-energy can be written as
$\Sigma_q(U)=z_0^2 \Sigma_q(U=0)$ so that upon approaching half-filling
the self-energy is determined by both the diverging $\Sigma_q(U=0)$
and the Gutzwiller renormalization factor $z_0^2$ which at $n=1$ tends to
zero for $U>U_{c}$ (in Fig. \ref{phase} the solid bar at $n=1$ indicates this
regime where the charge carriers are  localized). 
As a consequence $|\Sigma_q(U)|$ develops a maximum 
as a function of concentration and fixed $U$ which is reflected in the
maximum of the instability line.

Another instability occurs when the system is stable against nesting
(i.e. $|\Sigma_{q=2k_F}| < \Omega_{q=2k_F}/2$) but the slope of 
$|\Sigma_{q\to 0}|$ becomes larger than the slope of  $\Omega_{q\to 0}$.
Then there exists another solution of the condition Eq. (\ref{eq:cond})
The transition towards this instability occurs at $q=0$ when both
slopes become equal. The corresponding line is shown in Fig. \ref{phase})
by the dashed dotted curve.  Similar to the Holstein-Hubbard model we
thus find a $q=0$ instability for large $U$ which here is confined
to a region close to half-filling.
However, in contrast to the Holstein model, where the
phase separation is due to an instability of the 'zero sound'
particle-hole collective mode $v_{\rho}q$ caused by the coupling
to the optical phonon we have in the SSH model a coupling
between two acoustic modes, i.e. $\Omega_q\approx v_{ph}q$ and 
$v_{\rho} q$. 
For the situation we have analyzed in Fig. \ref{phase} we find 
always $v_{ph}<v_{\rho}$  so that the mode which becomes
unstable has dominantly phonon character.
However, since $v_{\rho}$ in the Mott regime is
renormalized to very small values one could also
imagine a situation where $v_{ph}>v_{\rho}$ in analogy to the
Holstein case. 

Finally, we have seen that the TDGA applied to the SSH coupling yields
an effective phonon dispersion Eq. (\ref{eq:omeff}) which yields
correlation induced softening through the elimination of
double occupancy fluctuations (cf. Fig. \ref{deltaomega}).
For large coupling $\tilde{g}$ these frequencies can become negative
even without the consideration of density fluctuations.
The corresponding regime in Fig. \ref{phase} is enclosed  by the dashed
line. We find (at least for the present model) 
that this area is always in a parameter regime
which corresponds to the nesting induced instability and therefore
never gives rise to a 'real' instability.

\section{Conclusions}
We have investigated the renormalization of phonon frequencies
within the Hubbard-Holstein and Hubbard-SSH models based on
the TDGA approach. 
Our considerations of the Holstein coupling for one-dimensional
correlated systems supplements our investigations in higher
dimensions \cite{diciolo08} and serves as a reference
for our computations of the SSH coupling. In the latter
case we have found that correlations influence on the
phonon modes $\Omega_q$ via two mechanisms. First, the coupling to
double occupancy fluctuations leads to a softening which has a
maximum around $q=\pi/2$, depending on $U$ and doping.
The second mechanism is the standard screening from density
fluctuations. However, in this regard our TDGA approach goes beyond 
the standard RPA since it incorporates the interaction between 
phonons and both,
transitive and induced local density fluctuations. 
This leads to an interesting dependence of the self-energy
$\Sigma_q$ on the local repulsion $U$ since it becomes suppressed
for large but enhanced for small momenta.

We have found that also the transitive coupling of the Hubbard-SSH model
gives rise to an interesting phase diagram where correlations
can suppress the $q=2k_F$ nesting instability but at the same time
are responsible for the occurence of a $q=0$ instability in the
vicinity of half-filling.
Our calculations therefore indicate that the phase separation instability,
previously only evidenced for Holstein-type couplings, seems to be a generic 
property of strongly correlated electrons coupled to phonons.
This is interesting in the context of complex oxides which often show
nanoscale phase separation. We have restricted to a short range only
model, however the short range phase when supplemented with the 
 long range Coulomb interaction is well known to lead to mesoscopic inhomogeneities.~\cite{CDCG95,lor01I,lor02,low94,Ort06,Ort07,ort08}

How do our results apply for higher-dimensional systems, especially
with regard to the anomalous softening of bond-stretching modes
in perovskite materials?~\cite{queeney99,pintch99,Reich96,uchiyama04,DAs02,graf08}
Consider e.g. the half-breathing mode in cuprates which
involves the movement of two planar oxygen ions towards 
the central Cu ion. The induced change of the ionic
potential on Cu leads to a Holstein-type coupling whereas
the associated modulation of the Cu-O hopping integral gives
rise to a coupling of the SSH type.
Concerning the latter interaction it is interesting that the
double occupancy induced renormalization in the 2-D three-band model
would lead to a maximum
frequency shift at the zone boundary (in contrast to that at $q=\pi/2$ in
the 1D SHH model). This kind of interaction therefore induces a downwards
dispersion of the half-breathing
mode which in the lowest approximation  just vibrates at constant frequency.
Obviously, in order to account for the doping dependence of the 
softening one has additionally to consider the effect from the
density fluctuations entering the phonon self-energy. In this
regard it would be interesting to investigate wether our approach
can improve related Hartree-Fock (HF) calculations within the three-band model \cite{roesch04} which give an incorrect (i.e. too small) 
doping dependence of the softening. In fact, since the correlation
functions in the TDGA incorporate the correlation induced reduction of
the kinetic energy its dependence on the charge carrier concentration
is expected to be much more pronounced than in the HF approach.
It should be noted that calculations of the
density response for the  $tJ$-model also indicate a strong
renormalization of bond-stretching phonons \cite{khali96}
with a larger anomaly occuring for half-breathing as compared
to full-breathing modes.~\cite{zhang07}

Our theory can be easily extended towards ground states which break
translational symmetry. In this regard it would be interesting to
evaluate the phonon renormalization from striped ground states
since there is experimental evidence \cite{Rez06,Rez07} that these textures 
contribute to the anomalous phonon softening 
at intermediate $q$ values in high-temperature superconductors. 
Since codoped LSCO compounds,  where  
static stripe order is unambigously established, show a rather strong 
renormalization it has been
argued \cite{mukhin07} that the corresponding phonon dispersion 
exhibits a Kohn-type anomaly
originating from the $q=2k_F=\pi/2$ scattering along the half-filled
stripes. Since the GA (in contrast to HF) leads to 
half-filled stripes as stable mean-field solutions of the
Hubbard model \cite{lor02b} our approach allows for a test of 
this scenario from a realistic model.

An interesting issue concerns the question wether the correlation induced 
enhancement of the phonon self-energy for small wave-vectors in
case of the SSH coupling also reflects a corresponding enhancement
of the electron-phonon vertex and eventually superconducting correlations.
In the present model superconductivity is mediated by the transitive lattice
fluctuations which, as we have shown, are not screened in the
same way as the local ones and in the long wavelength limit even can be
underscreened. This scenario thus appears similar in spirit to previous works
by Capone and coworkers \cite{cap2001,cap2004} where the enhancement of
Cooper-pairing near the Mott transition of multiband Hubbard models was
investigated. The interorbital fluctuations 
in these models are only little affected by the correlations (which however 
lead to a decrease  of the quasiparticles bandwidth and a concomitant increase 
of the density of states near the Fermi level) so that they can effectively 
enhance superconductivity. An analogous mechanism may also work in the
correlated SSH model.
However, this issue is much more involved since the investigation of
pair-pair scattering requires
a GA energy functional which is charge-rotationally invariant \cite{sei08}
also for the transitive electron-phonon coupling Eq. (\ref{eq:ssh2}).
As a consequence the coupling between pair and lattice fluctuations
will in general be different from ${\bf W_q}^{el-ph}$ given in
Eqs. (\ref{eq:w1},\ref{eq:w2}) and will considered elsewhere.

\acknowledgments
We are grateful to Bob Markiewicz for enlightenment  comments. 
E.v.O., M.G., J.L. and G.S. acknowledge financial support from the 
Vigoni foundation.

\section*{Appendix}
In the TDGA expansion Eq.\ref{eskspace} we have introduced the 
following abbreviations for the $z$-factors and its derivatives:

\begin{eqnarray*}
 && z_{i\sigma}\equiv z_0 , \ \ \frac{\partial z_{i \sigma}}{\partial \rho_{ii 
\sigma}}\equiv z^{'},  \nonumber \\ &&  \frac{\partial z_{i\sigma}}
{\partial \rho_{ii -\sigma}}\equiv
 z^{'}_{+-}, \
\frac{\partial z_{i \sigma}}{\partial D_{i}}\equiv z^{'}_{D}
\label{z1} \\
&& \frac{\partial^2 z_{i \sigma}}{\partial \rho^{2}_{ii \sigma}} 
\equiv z^{''}_{++}, \
\frac{\partial^2 z_{i \sigma}}{\partial \rho_{ii \sigma}\partial \rho_{ii -\sigma}}
\equiv z^{''}_{+-}, \ \frac{\partial^2 z_{i \sigma}}{\partial 
\rho^{2}_{ii -\sigma}} \equiv z^{''}_{--}  \nonumber \\
&&  \frac{\partial^2 z_{i \sigma}}{\partial D^{2}_{i}}\equiv z^{''}_{D}, \ 
\frac{\partial^2
z_{i \sigma}}{\partial \rho_{ii \sigma}\partial D_{i}}\equiv z^{''}_{+D}, 
\ \frac{\partial^2
z_{i \sigma}}{\partial \rho_{ii -\sigma}\partial D_{i}}\equiv z^{''}_{-D}
\label{z2}
\end{eqnarray*}
For the half-filled paramagnetic state we have  $z^{'} = z^{'}_{+-}$
and $z^{''}_{+D}=z^{''}_{-D}$.

 \end{document}